\documentclass[twocolumn]{aastex63}

\usepackage{epsfig,amsmath,amsfonts,amssymb,graphicx,subfigure,enumitem}
\usepackage[english]{babel}
\usepackage{blindtext}
\usepackage{tikz}
\usepackage{todonotes}
\usepackage{float}
\usepackage{booktabs}
\usepackage{gensymb}
\usepackage{float}
\usepackage{newtxtext,newtxmath}
\usepackage[T1]{fontenc}
\usepackage{ae,aecompl}
\usepackage{graphicx}	
\usepackage{amsmath}	
\usepackage{amssymb}
\usepackage{multirow}
\usepackage{graphicx}
\usepackage{float}
\usepackage{subfloat}

\newsavebox\mybox
\newsavebox{\tempbox}
\hypersetup{linkcolor=magenta,citecolor=blue,filecolor=cyan}

\newsavebox{\leftbox}
\newsavebox{\rightbox}

\newcommand{\s}{${\omega}$\,} % Omega
\def\m{\textminus} % counts s^-1

\shorttitle{Optical Studies of RBS 0490 and SDSS J075939.79+191417.3}
\shortauthors{Joshi et al.}

\begin{document}

\title{Optical characterization of two cataclysmic variables: RBS 0490 and SDSS J075939.79+191417.3}
\author[0000-0003-2431-981X]{Arti Joshi}
\affiliation{School of Physics and Technology, Wuhan University, Wuhan 430072, China}
\author{J.C.Pandey}
\affiliation{Aryabhatta Research Institute of Observational Sciences (ARIES), Nainital - 263002, India}
\author{Nikita Rawat}
\affil{Aryabhatta Research Institute of Observational Sciences (ARIES), Nainital - 263002, India}
\author{Ashish Raj}
\affil{Department of Physics and Astrophysics, University of Delhi, 110007 Delhi, India}
\author{Wei Wang}\thanks{Corresponding author\\ wangwei2017@whu.edu.cn}
\affiliation{School of Physics and Technology, Wuhan University, Wuhan 430072, China}
\affiliation{WHU-NAOC Joint Center for Astronomy, Wuhan University, Wuhan 430072, China} 
\author{H. P. Singh}
\affil{Department of Physics and Astrophysics, University of Delhi, 110007 Delhi, India}
\label{firstpage}

\renewcommand{\abstractname}{ABSTRACT}
\begin{abstract} 
We present optical photometric and spectroscopic observations of two Cataclysmic Variables (CVs), namely RBS 0490 and SDSS J075939.79+191417.3. The optical variations of RBS 0490 have been found to occur at the period of 1.689$\pm$0.001 hr which appears to be a probable orbital period of the system. Present photometric observations of SDSS J075939.79+191417.3 confirm and refine the previously determined orbital period as 3.14240928$\pm$0.00000096 hr. The presence of long-duration eclipse features in the light curves of SDSS J075939.79+191417.3 indicates eclipses might be due to an accretion disc and bright spot. The orbital inclination of SDSS J075939.79+191417.3 is estimated to be $\sim$ 78\hbox {$^\circ$} using the eclipse morphology. The phased-light curve variations during the orbital cycle of RBS 0490 provide evidence of the emission from an independent second accretion region or a second fainter pole. Optical spectra of RBS 0490 and SDSS J075939.79+191417.3 show the presence of strong Balmer, weak He II ($\lambda$4686) emission lines, along with the detection of strong $H\beta$ emission lines with a large value of equivalent width. The characteristic features of RBS 0490 seem to favour low-field polars, while SDSS J075939.79+191417.3 appears to be similar to the non-magnetic systems. 
\end{abstract}


%%%%%%%%%%%%%%%%%%%%%%%%%%%%%%%%%%%%%%%%%%%%%%%%%%%%%%%%%%%%%%%%%%%% Introduction %%%%%%%%%%%%%%%%%%%%%%%%%%%%%%%%%%%%%%%%%%%%%%%%%%%%%%%%%%

\section{Introduction}
\label{sec:intro}
Cataclysmic Variables are semi-detached close binary systems in which white dwarf (WD) primary accretes material from the mass donating secondary via Roche-lobe overflow \citep{Warner95}. The material transferred from the secondary flows through the inner Lagrangian point and orbits around the primary, forming an accretion disc in non-magnetic systems. However, if the magnetic field strength of the WD is large enough, an accretion disc can form far from the WD, but the inner disc disrupts at the magnetospheric radius, and the field forces material to fall onto the poles of the WD. In these systems, $P_\omega$ $<$ $P_\Omega$, where $P_\omega$ and $P_\Omega$ are spin and orbital periods, respectively, and are known as Intermediate Polars (IPs) \citep[][ and references therein]{Patterson94}. Alternately, if the magnetic field of the WD is sufficiently strong, the magnetic pressure exceeds the ram pressure, preventing the formation of an accretion disc. The accreted matter follows the magnetic field lines of the WD and dumps onto the poles \citep[see][]{Cropper90, Warner95}. These binary systems are  known as polars for which $P_\omega$ = $P_\Omega$. The majority of polars have orbital periods shorter than the `period gap of 2-3 hr' \citep{Scaringi10}. The optical and near-infrared radiations in these systems are dominated by cyclotron emission and originate from the accretion column near the WD in the post-shock region. These systems tend to show periodic modulation at the orbital period, and their optical spectra exhibit strong Balmer emission lines along with He I and He II emission lines \citep[][and references therein]{Warner86}. The H$\beta$ and HeII ($\lambda$4686 \AA) emission lines are much stronger in polars and usually originate in the accretion column. The majority of the polars are single-pole accretors \citep[e.g.][]{Schwope93, Salvi02, Bridge03}, whereas some polars accrete onto both poles \citep[e.g.][]{Beardmore95, Schwope95, Schmidt99} and some other switch between single and double pole accretion \citep[e.g.,][]{Rosen96, Mason98}. In this paper, we present detailed analyses of two CVs namely RBS 0490 and SDSS J075939.79+191417.3 (hereafter J0759) using the optical photometric and spectroscopic observations that were taken between the years 2005 and 2021.

%-------------------------------------------------------------------
\begin{deluxetable*}{lclllcrccc}
\tablecolumns{8}
\tablewidth{-0pt}
\tabletypesize{\normalsize}
\setlength{\tabcolsep}{0.05in}
  \tablecaption{\leftskip0mm{Log of optical photometric and spectroscopic observations of RBS 0490 and J0759. \label{tab:obslog}}}
\tablehead{
  {\textbf {Object}} &{\textbf {Date of}} &{\textbf {~~~~Facility}}&{\textbf {~~~~~Instrument}}&{\textbf {~~~~Filter/}}&{\textbf {Integration}}    & \colhead{~~{\textbf {Time}}}        &{\textbf {Time-span}} \\
       & {\textbf {~~Observations}} &         &          &{\textbf {~~~~Band}}   &{\textbf {Time (s)}} & {\textbf {(JD$_{t}$)}} &  {\textbf {(hr)}}    &     }
\startdata                                                                         
  {\bf  RBS 0490}              & 2015 Oct 10		   	  & 1.04 m - ST     &\colhead{1k$\times$1k CCD}        &\colhead{R}   & \colhead{300}   &2457305.00 &\colhead{3.52}\\ 
			       & 2015 Nov 06			  & 1.3 m - DFOT    &\colhead{512$\times$512 CCD}      &\colhead{R}   & \colhead{200}   &2457333.00 &\colhead{3.87}\\  
			       & 2017 Dec  22			  & 2.01 m - HCT    &\colhead{HFOSC/Gr7}              &380-780 nm    & \colhead{3600}  &2458110.08 &\colhead{1.00}\\
			       & 2018 Nov  30			  & 2.01 m - HCT    &\colhead{HFOSC/Gr7}              &380-780 nm    & \colhead{3000}  &2458453.27 &\colhead{0.83}\\
{\bf J0759}                    & 2016 Mar  07			  & 1.3 m - DFOT    &\colhead{512$\times$512 CCD}     &\colhead{R}   & \colhead{300}   &2457455.00 &\colhead{4.11}\\
			       & 2021 Feb  06		          & 1.3 m - DFOT    &\colhead{2k$\times$2k CCD}       &\colhead{R}   & \colhead{300}   &2459252.00 &\colhead{3.49}\\
                               & 2021 Feb  07			  & 1.3 m - DFOT    &\colhead{2k$\times$2k CCD}       &\colhead{R}   & \colhead{300}   &2459253.00 &\colhead{3.40}\\
                               & 2021 Feb  08 			  & 1.3 m - DFOT    &\colhead{2k$\times$2k CCD}       &\colhead{R}   & \colhead{300}   &2459254.00 &\colhead{3.52}\\ 
                               & 2021 Feb  15			  & 1.04 m - ST     &\colhead{4k$\times$4k CCD}       &\colhead{R}   & \colhead{300}   &2459261.00 &\colhead{3.24}\\ 
                               & 2018 Dec  17			  & 2.01 m - HCT    &\colhead{HFOSC/Gr7}              &380-780 nm    & \colhead{3600}  &2458470.39 &\colhead{1.00}\\
                               & 2019 Feb  09			  & 2.01 m - HCT    &\colhead{HFOSC/Gr7} 	      &380-780 nm    & \colhead{3600}  &2458524.36 &\colhead{1.00}\\
                               & 2020 Jan  14			  & 2.01 m - HCT    &\colhead{HFOSC/Gr7} 	      &380-780 nm    & \colhead{2700}  &2458863.45 &\colhead{0.75}\\
                               & 2021 Mar  31    		  & 2.01 m - HCT    &\colhead{HFOSC/Gr7} 	      &380-780 nm    & \colhead{3600}  &2459305.08 &\colhead{1.00}\\
			       & 2021 Apr  24			  & 2.01 m - HCT    &\colhead{HFOSC/Gr7} 	      &380-780 nm    & \colhead{3600}  &2459329.10 &\colhead{1.00}\\
\enddata
\tablecomments 
{JD$_{t}$ for HFOSC is the start time of the observations, whereas for photometric observation it is observation day. } \end{deluxetable*}
%-------------------------------------------------------------------

RBS 0490 was detected during the {\it ROSAT} all-sky survey \citep[RASS;][]{Voges99, Schwope02} and found to be {\it ROSAT} bright source. Inverting the Gaia parallax of RBS 0490 gives a distance of 320$\pm$11 pc \citep{GaiaCollaboration21}. \citet{Schwope02} constructed a spectrum from 87 photons detected in the RASS and suggested that it can be modeled with a thermal bremsstrahlung component at a temperature of 20 keV. All emission lines appeared double-peaked in the optical spectrum of \citet{Schwope02}. Moreover, H-Balmer and He I emission lines were very prominent, whereas He II emission lines were relatively weak. Later, \citet{Thorstensen06} found a candidate 46-min radial-velocity period. In contrast to \citet{Schwope02}, \citet{Thorstensen06} found single-peaked emission lines, and only moderately strong He II $\lambda$4686. \cite{Harrison15} have analysed the infrared data obtained from the Wide-field Infrared Survey Explorer (WISE) bands and suggested that it is a polar.

J0759 was identified as a CV by \cite{Szkody06} based on the optical spectrum obtained during the Sloan Digital Sky Survey (SDSS). The SDSS spectrum shows strong Balmer emission (H$\beta$ has an equivalent width (EW) of 72 \AA), and a He II($\lambda$4686)/H$\beta$ EW ratio of 0.2. Subsequently, they reported that J0759 reveals the strength of He II ($\lambda$4686 \AA) larger than most dwarf novae but not quite up to the values typical for polars or IPs. Later, \cite{Gansicke09} reported the orbital period of $\sim$ 188.45 min for J0759 using the SDSS spectroscopy. However, a much refined period of 0.1309337 d is reported in the AAVSO VSX catalog based on the data obtained from the Catalina Real-time Transit Survey (CRTS) and Zwicky Transient Facility (ZTF). Using the Gaia parallax \citep{GaiaCollaboration21}, the distance of J0759 is calculated to be 1873$\pm$535 pc.

The paper is organized as follows. We summarize the description of observations and data reduction in the next section. Analyses and the results of the optical data are described in section \ref{sec:ana}. Finally, we present discussion and conclusions in sections \ref{sec:diss} and \ref{sec:conc}, respectively.

%-------------------------------------------------------------------
\begin{deluxetable}{lcccrcrcccrcccr}
\tablecolumns{9}
\tablewidth{-0pt}
\tabletypesize{\normalsize}
\setlength{\tabcolsep}{0.05in}
\tablecaption{\leftskip0mm{Comparison stars used for the differential photometry for RBS 0490 and J0759.\label{tab:obj_comp}}}
\tablehead{
  {\textbf {Object}} & {\textbf {Reference}}  &\colhead{{\textbf {B1}}}&\colhead{{\textbf {R1}}} &\colhead{{\textbf {B2}}} & \colhead{{\textbf {R2}}} \\
       & {\textbf {USNO-B1.0}}  &{\textbf {(mag)}}&{\textbf {(mag)}}&{\textbf {(mag)}}&{\textbf {(mag)}} }
\startdata                                                                         
  {\bf RBS 0490}                    & 0731-0063800 & 16.89    & 16.53 & 18.30 & 17.11 \\
`C1'		                    & 0731-0063769 & 15.20    & 13.96 & 15.26 & 14.12 \\
`C2'                                & 0731-0062447 & 16.63    & 14.66 & 16.95 & 15.40 \\ 
  {\bf J0759}                       & 1092-0155844 & 18.26    & 18.50 & 18.65 & 17.84   \\  
  `C1'	    	                    & 1092-0155866 & 16.81    & 15.78 & 16.79 & 15.63   \\
  `C2'                              & 1092-0155859 & 17.13    & 15.73 & 16.94 & 15.65   \\ 
\enddata
\tablecomments{C1 and C2 stands for comparison 1 and 2 for each program star.}
\end{deluxetable}
%-------------------------------------------------------------------

%%%%%%%%%%%%%%%%%%%%%%%%%%%%%%%%%%%%%%%%%%%%%%%%%%%%%%%%%%%%%%%%%%%%%%%%%%%%% Observations And Data Reduction %%%%%%%%%%%%%%%%%%%%%%%%%%%%%%%%%%%%%%%%%%%%%%%%%%%%%%

\section{Observations And Data Reduction}
\label{sec:obs_red}
\subsection{Optical Photometric Observations} 
\label{sec:obs_phot}
R-band photometric observations of both sources were carried out during epochs 2015, 2016, and 2021 using the 1.04-m Sampuranand Telescope \citep[ST;][]{Sinvhal72} and 1.3-m Devasthal Fast Optical Telescope  \citep[DFOT;][]{Sagar11} of ARIES, Nainital, India. A detailed log for photometric observations is given in Table \ref{tab:obslog}. The ST was equipped with a 1k$\times$1k CCD and 4k$\times$4k CCD during observations of epochs 2015 and 2021, respectively. However, the DFOT was equipped with CCDs 2k$\times$2k during the epoch 2021 and 512 $\times$ 512 during observations of epochs 2015 and 2016. Several bias and twilight sky flat frames were also acquired during each observing run. Pre-processing of each observed frame was performed using IRAF\footnote{\textcolor{magenta}{IRAF is distributed by the national optical astronomy observatories, USA.}} software. To obtain the instrumental magnitudes, the differential photometry in a manner variable minus comparison star was performed by using a comparison star in the same field. The details of the adopted comparison stars labeled as `C1' and `C2' for each source are given in Table \ref{tab:obj_comp}. The constant brightness of comparison stars was checked by inspecting the light curves of `C1-C2' for each source separately. For consecutive epoch of observations (see Table \ref{tab:obslog}), the nightly mean of the standard deviation of `C1-C2' were derived as 0.0052 and  0.0045  for RBS 0490, and 0.0055, 0.0024, 0.0040, 0.0040, and 0.0091 for J0759. Finally, the R-band magnitudes of observations were computed with respect to comparison star `C1'. 

\subsection{TESS, ZTF, and CRTS Observations} 
\label{sec:obs_tess}
RBS 0490 was observed by the Transiting Exoplanet Survey Satellite (TESS) with camera 2 in sector 31 for 25 days at a two-minute cadence from 22 October 2020 to 16 November 2020. TESS was launched on 18 April 2019 into a 13.7 d orbit. It has four small telescopes with four cameras with field of view of each 24$\times$24 degree$^2$ are aligned to cover 24 $\times$ 90-degree strips of the sky called `sectors' \citep[see][for details]{Ricker15}. TESS bandpass extends from 600 to 1000 nm with an effective wavelength of 800 nm. TESS observations of RBS 0490 were continuous except for $\sim$ 56.8-hour gap halfway into the sector. The data of RBS 0490 was stored under Mikulski Archive for Space Telescopes (MAST) data archive with identification number `TIC 279483438'. Data taken during an anomalous event had quality flags greater than 0 in the FITS file data structure and thus we have considered only the data with the `QUALITY flag' = 0. We have taken PDCSAP flux values, which is the Simple Aperture Photometry flux after they have been corrected for systematic trends common to all-stars.  

The ZTF is a northern-sky synoptic optical survey for high cadence time-domain astronomy, which used the Palomar 48-inch (P48) telescope \citep{Graham19, Bellm19b, Dekany20}. Its CCD camera consists of 16 6K$\times$6K E2V CCDs, producing a 47 deg$^2$ field of view. There are three available filters, ZTF-g, ZTF-r, and ZTF-i. Exposure times are usually 30 seconds for g and r bands and 60 seconds for i band. RBS 0490 and J0759 were observed by ZTF between 2018 to 2021. ZTF lists a total of 339 exposures for RBS 0490 from 23 September 2018 to 26 February 2021 and 520 exposures for J0759 from 28 March 2018 to 24 April 2021. 

J0759 was also observed by the CRTS \citep{Drake09, Drake14} between 09 April 2005 to 24 December 2013 in the V-filter. CRTS combines three distinct surveys$-$the Mount Lemmon Survey (MLS) and the Catalina Schmidt Survey (CSS) in the Northern Hemisphere and the Siding Spring Survey (SSS) in the South. The CRTS data of J0759 is available from the MLS and CSS surveys, and its light curves are easily accessible online.

%==========================
\begin{figure*}[ht]
\centering
  \subfigure[]{\includegraphics[width=165mm, height=45mm]{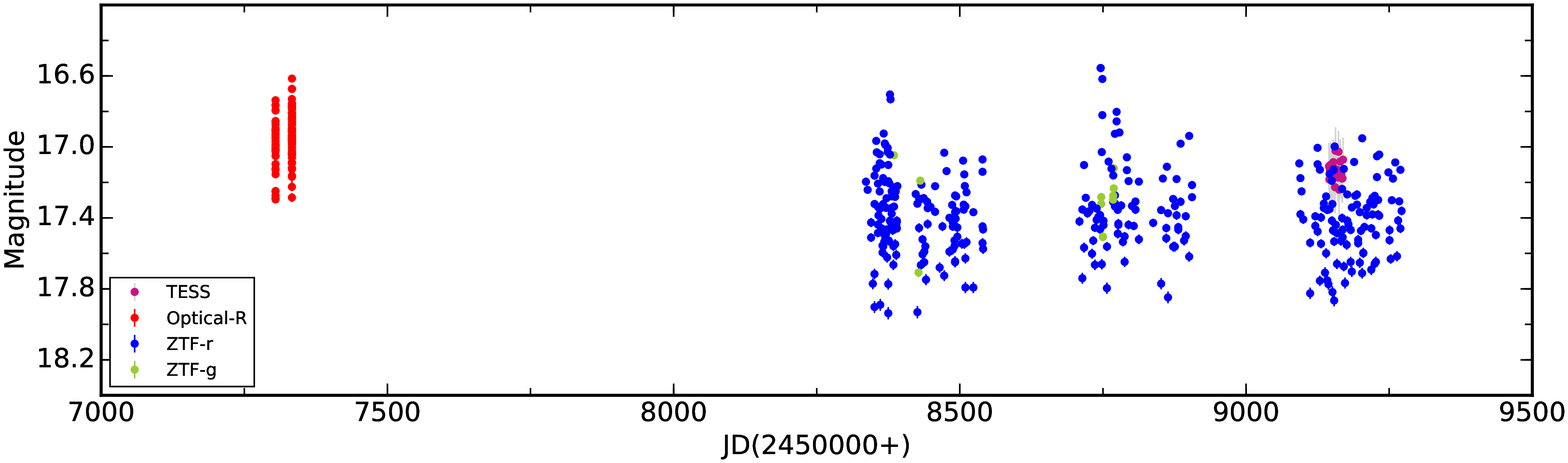}\label{fig:longterm_lc_RBS490}}
  \subfigure[]{\includegraphics[width=165mm, height=45mm]{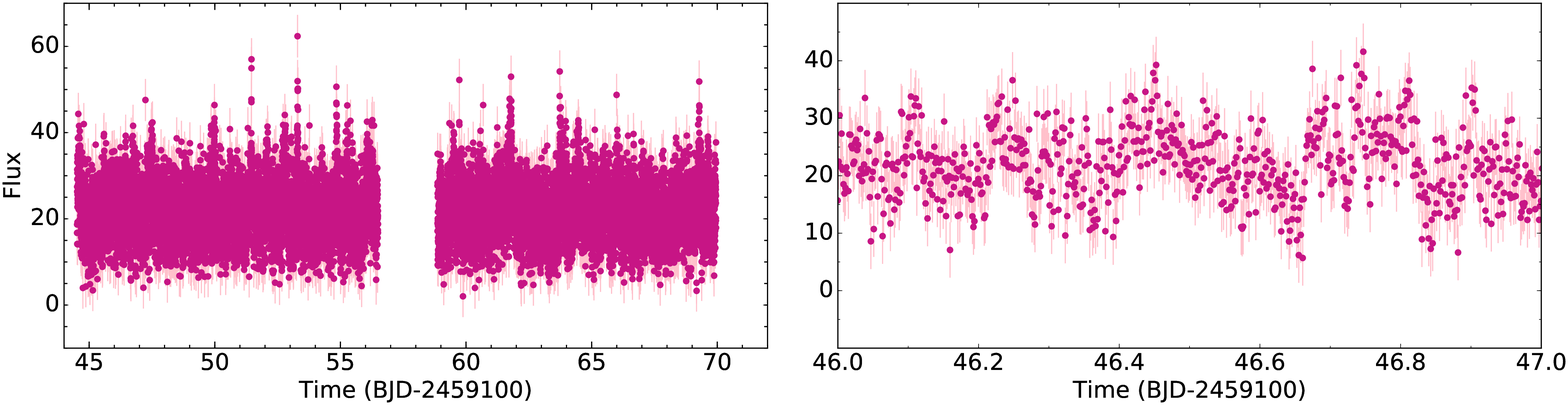}\label{fig:tess_lc_RBS490}}
  \subfigure[]{\includegraphics[width=165mm, height=45mm]{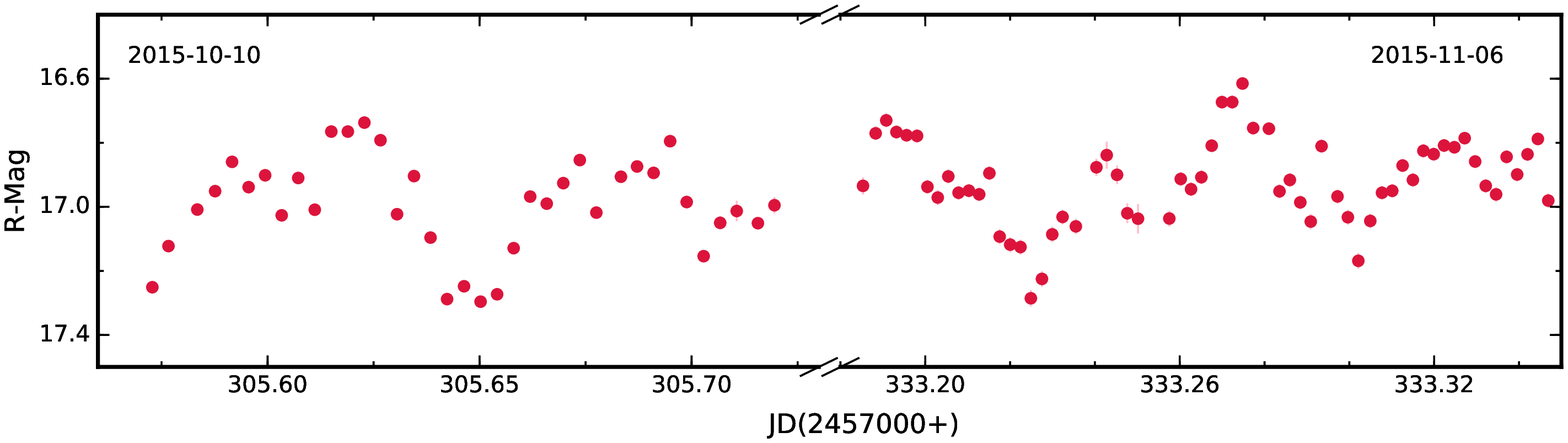}\label{fig:opt_lc_RBS490}}
  \caption{(a) Combined R-band, TESS, and ZTF-r/g light curves of RBS 0490, (b) the full TESS light curve of RBS 0490, where the right panel shows zoomed version of the one-day observation, and (c) R-band light curves of RBS 0490 for two epochs of observations.}
\end{figure*}
%==========================
%==========================
\begin{figure*}
\centering
\includegraphics[width=95mm, height=110mm]{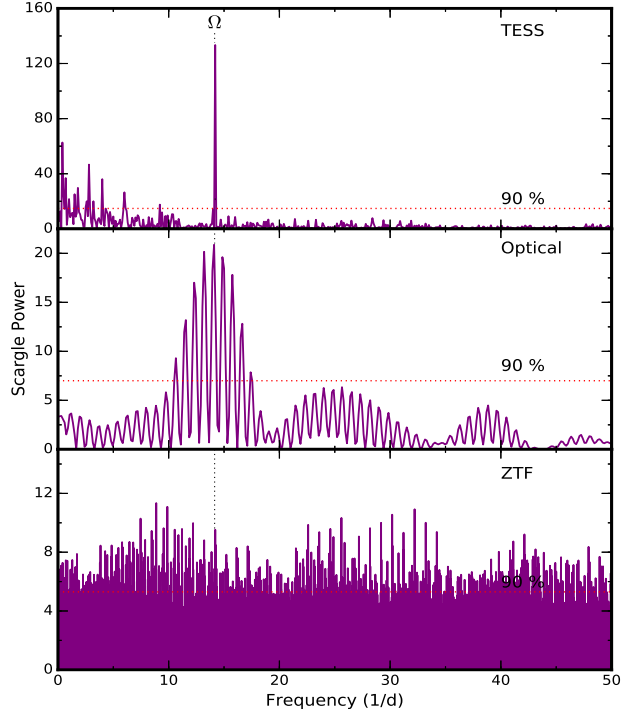}
\caption{The Lomb-Scargle power spectra of RBS 0490 obtained from the complete data set of the TESS, R-band, and ZTF-r band observations. The horizontal dotted lines in the Lomb-Scargle power spectra represent the 90 \% confidence level. } 
\label{fig:ps_RBS490}
\end{figure*}
%==========================

%-------------------------------------------------------------------
\begin{table*}
\normalsize
\begin{center}
\caption{Periods corresponding to dominant peaks in the power spectra of RBS 0490 and J0759 obtained from the R-band photometry, TESS, ZTF, and CRTS  observations.\label{tab:ps}}
\setlength{\tabcolsep}{0.05in}
  \begin{tabular}{cccccccccccc}
\tableline\tableline
\tableline 
 Period(hr) 	 &\multicolumn{1}{c}{\textsc{R-band photometry}}   && \multicolumn{1}{c}{\textsc {TESS/CRTS*}} &&   \multicolumn{1}{c}{\textsc {ZTF}}  \\
 \hline
					       && \multicolumn{3}{c}{\textsc{ {\bf RBS 0490}}}                           &&         \\
{\bf $P$$_\Omega$}     	 & 1.6894$\pm$0.0011                                 && 1.6894$\pm$0.0012   && 1.6895$\pm$0.0003             \\
    		 & & \multicolumn{3}{c}{\textsc{ {\bf J0759}}}           &         \\
 {\bf $P$$_\Omega$ }      & 3.149025$\pm$0.011338      && 3.142396$\pm$0.000032 && 3.142387$\pm$0.000092                     \\
 {\bf $P_2$$_\Omega$}      & 1.573160$\pm$0.002824      && 1.571206$\pm$0.000008 && 1.571194$\pm$0.000023                    \\
 {\bf $P_3$$_\Omega$}      & 1.054452$\pm$0.001268      && 1.047469$\pm$0.000004 && 1.047473$\pm$0.000010                    \\
 {\bf $P_4$$_\Omega$}      & 0.789414$\pm$0.000711      && 0.789678$\pm$0.000002 && 0.785603$\pm$0.000006                   \\ 
 \tableline
\end{tabular}
\end{center}
*Period obtained from TESS data for RBS 0490 and CRTS data for J0759. 
\end{table*}
%-------------------------------------------------------------------

%==========================
\begin{figure*}
\centering
\includegraphics[width=130mm, height=110mm]{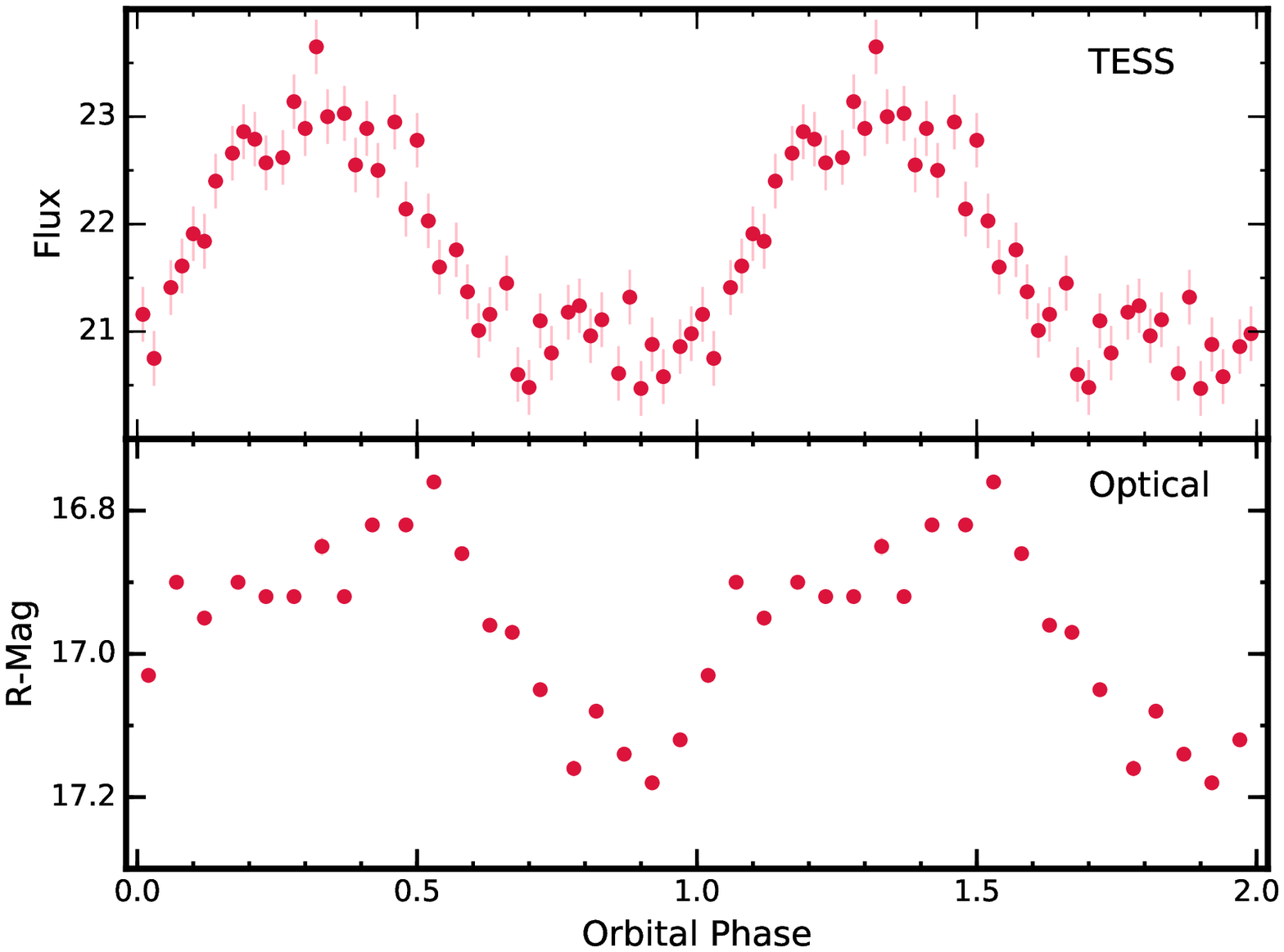}
\caption{Orbital-phase-folded light curves of RBS 0490 obtained from the TESS and R-band observations with phase bin of 0.0222 and 0.05, respectively.} 
\label{fig:tess_opt_flc_RBS490}
\end{figure*}
%==========================

\subsection{Optical Spectroscopic Observations} 
\label{sec:obs_spec}
We obtained long-slit low-resolution optical spectra of both sources between the year 2017 to 2021 using the 2.01-m Himalayan Chandra Telescope at IAO, Hanle which is equipped with the Hanle Faint Object Spectrograph and Camera \citep[HFOSC;][]{Prabhu10}. A detailed log of the spectroscopic observations is given in Table \ref{tab:obslog}. For all epochs, the observations were taken using the grism Gr7 (3800-7800) \AA ~with a resolution of 1330 and dispersion of 1.45 \AA/pixel. Spectrophotometric standard stars were also observed during each observing run. The data reduction of the standard and program stars was accomplished with the three main steps using IRAF tasks: pre-processing, wavelength calibration, and flux calibration. Lamp spectra of Fe-Ar were used for the wavelength calibration. The night sky emission lines 5577 \AA, 6300 \AA, and 6363 \AA ~were used to check the appropriate shift, and the shift was then applied with {\sc specshift} task wherever it was required. 

%%%%%%%%%%%%%%%%%%%%%%%%%%%%%%%%%%%%%%%%%%%%%%%%%%%%%%%%%%%%%%%%%%%%%%%%%%%%% Analysis and Results %%%%%%%%%%%%%%%%%%%%%%%%%%%%%%%%%%%%%%%%%%%%%%%%%%%%%%%%%%%%%%%%%

%%%%%%%%%%%%%%%%%%%%%%%%%%%%%%%%%%%%%%%%%%%%%%%%%%%%%%%%%%%%%%%%%%%%%%%%%%%%% RBS 0490 %%%%%%%%%%%%%%%%%%%%%%%%%%%%%%%%%%%%%%%%%%%%%%%%%%%%%%%%%%%%%%%%%%%%%%%%%%%%%
 
\section{Analysis and Results}
\label{sec:ana}
\subsection{RBS 0490}
\label{sec:ana_res_RBS490}
\subsubsection{Light Curve Morphology and Power Spectra}
\label{sec:lc_ps_RBS490}
Figure \ref{fig:longterm_lc_RBS490} combines the TESS, ZTF-r/g, and our own R-band observations of RBS 0490 from the year 2015 to 2021. To probe RBS 0490 in more detail, we have closely inspected its light curve observed from each facility independently. The full TESS light curve of RBS 0490 and zoomed version of its one-day observations are shown in Figure \ref{fig:tess_lc_RBS490}, whereas the R-band light curves are shown in Figure \ref{fig:opt_lc_RBS490}. Its light curve variations are associated mostly with the bright maxima along with minima and exhibit a periodic nature of the system. The periodic nature of the system has been demonstrated by applying the Lomb-Scargle (LS) periodogram algorithm \citep{Lomb76, Scargle82, Horne86} to the entire TESS, ZTF-r, and R-band observations. The LS power spectrum of the TESS data of RBS 0490 is shown in the top panel of Figure \ref{fig:ps_RBS490}. The observed highest significant peak in the TESS power spectrum corresponds to the period of 1.689$\pm$0.001 hr which is found to lie above the 99\% confidence level. The significance of this detected peak is determined by calculating the False Alarm Probability \citep[FAP;][]{Horne86}. The confidence level is represented by the dotted red lines. In contrast to radial-velocity measurements of \citet{Thorstensen06}, we have not detected any significant period near 46 min. We provisionally identify the 1.689-hr period with $P_\Omega$.

From the R-band photometric data, we have detected a significant period of 1.689$\pm$0.001 hr which is exactly similar to the period derived from the TESS data. The LS power spectrum obtained from the photometric data is shown in the middle panel of Figure \ref{fig:ps_RBS490}. The ZTF-r band light curve of RBS 0490 is very sparse and resulted in a noisy power spectrum as shown in the bottom panel of Figure \ref{fig:ps_RBS490}. The orbital frequency obtained from the ZTF data is also marked in this figure. The significant periods derived from the LS photometric, TESS, and ZTF data are given in Table \ref{tab:ps}.    

%==========================
\begin{figure*}
\centering
\includegraphics[width=130mm, height=100mm]{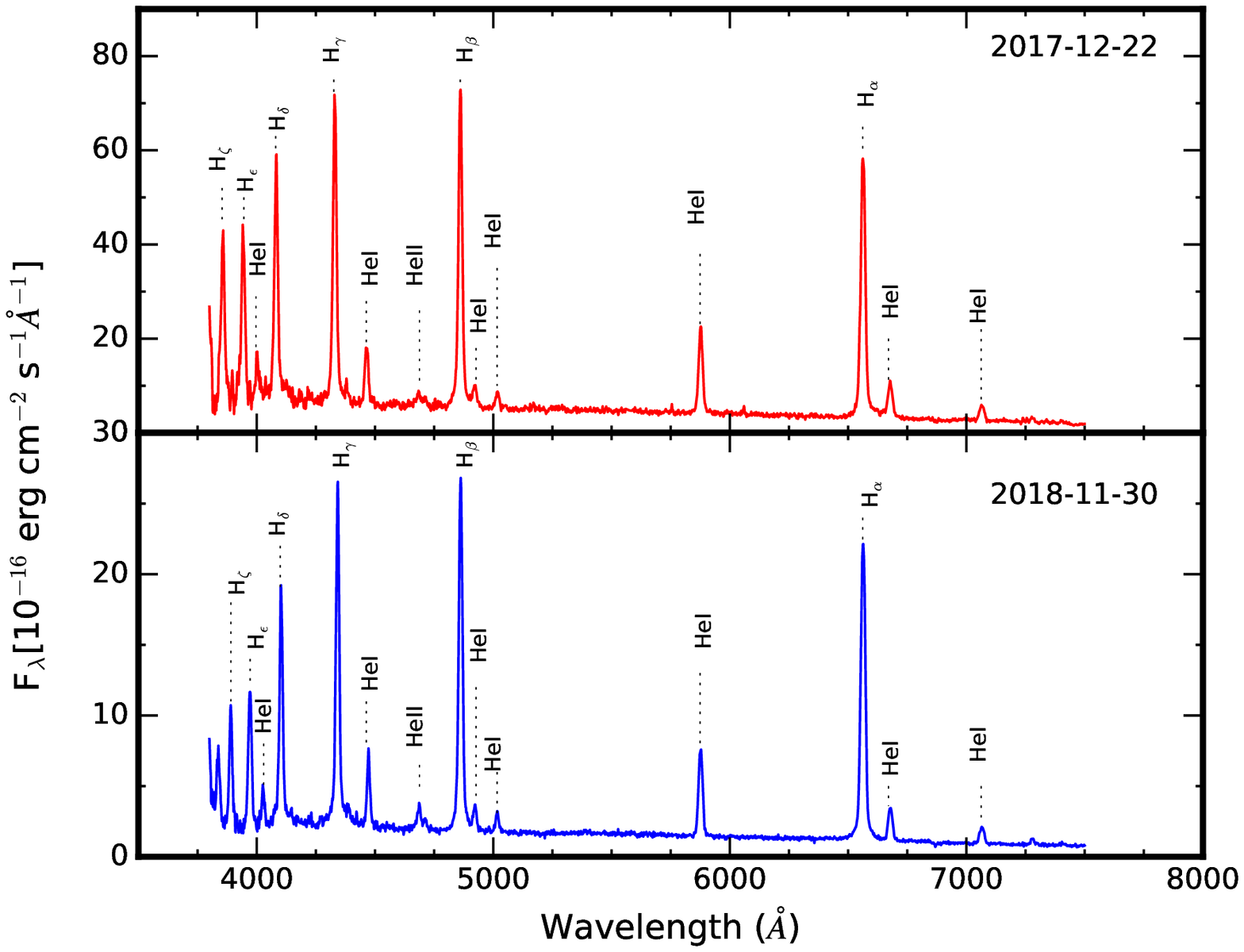}
\caption{Optical spectra of RBS 0490 for two epochs of observations. The epoch of the observation is mentioned in each panel of the spectrum.}
\label{fig:optspec_RBS490}
\end{figure*}
%==========================

%-------------------------------------------------------------------
\begin{table*}[ht]
\scriptsize
\centering 
\caption{Identification, flux, EW, and FWHM for emission features in the spectra of RBS 0490 and J0759.\label{tab:opt_spec_RBS490_J0759}}
\setlength{\tabcolsep}{0.025in}
\begin{tabular}{lccccccccccccccccccccccccccccccccccc}
\hline
\multirow{3}*{\textbf {Identification}}&& \multicolumn{7}{c}{\textbf{RBS 0490}}&&\multicolumn{19}{c}{\textbf{J0759}}\\
\cline{3-9}\cline{11-29}
&&\multicolumn{3}{c}{\textbf{2017-12-22}} && \multicolumn{3}{c}{\textbf{2018-11-30}}&&\multicolumn{3}{c} {\textbf{2018-12-17}}&&\multicolumn{3}{c}{\textbf{2019-02-09}} && \multicolumn{3}{c}{\textbf{2020-01-14}} &&\multicolumn{3}{c}{\textbf{2021-03-31}}        && \multicolumn{3}{c}{\textbf{2021-04-24}}\\
\cline{3-5} \cline{7-9} \cline{11-13} \cline{15-17} \cline{19-21} \cline{23-25} \cline{27-29}
&& \colhead{Flux}&\colhead{$-$EW}&\colhead{FWHM} &&\colhead{Flux}&\colhead{$-$EW}&\colhead{FWHM}&&\colhead{Flux}&\colhead{$-$EW}&\colhead{FWHM}&& \colhead{Flux}&\colhead{$-$EW}&\colhead{FWHM} &&\colhead{Flux}&\colhead{$-$EW}&\colhead{FWHM}&& \colhead{Flux}&\colhead{$-$EW}&\colhead{FWHM} &&\colhead{Flux}&\colhead{$-$EW}&\colhead{FWHM}\\
\hline
H$\zeta$ (3889~\AA)      && 126    & 54     &1123   &&705    &108    &1535   &&\nodata  &\nodata &\nodata &&\nodata&\nodata &\nodata&&\nodata &\nodata  &\nodata   &&\nodata  &\nodata &\nodata &&\nodata&\nodata  &\nodata\\
H$\epsilon$ (3970~\AA)   && 182    & 98     &1318   &&696    &109    &1390   &&\nodata  &\nodata &\nodata &&\nodata&\nodata &\nodata&&\nodata &\nodata  &\nodata   &&\nodata  &\nodata &\nodata &&\nodata&\nodata  &\nodata\\
HeI (4026~\AA)           && 40     & 18     &1094   &&150    &20     &1250   &&\nodata  &\nodata &\nodata &&\nodata&\nodata &\nodata&&\nodata &\nodata  &\nodata   &&\nodata  &\nodata &\nodata &&\nodata&\nodata  &\nodata\\
H$\delta$ (4102~\AA)     && 263    & 93     &1132   &&902    &101    &1265   && 41      & 31     &2774    &&32     &36      &1926   &&103     &92       &3583      && 50      & 30     &1535    &&72     &39       &2304  \\
H$\gamma$ (4340~\AA)     && 404    & 139    &1120   &&1170   &141    &1217   && 33      & 31     &2218    &&34     &26      &1576   &&70      &51       &2142      && 52      & 36     &1313    &&50     &31       &1832  \\
HeI (4471~\AA)	         && 84     & 40     &1032   &&213    &33     &1125   && 12      & 10     &2267    &&3      &3       &960    &&26      &24       &2281      && 15      & 14     &1140    &&9      &6        &1452  \\
CIII/NIII                &&\nodata &\nodata &\nodata&&\nodata&\nodata&\nodata&&\nodata  &\nodata &\nodata &&3      &3       &918    &&5       &4        &776       &&\nodata  &\nodata &\nodata &&14     &9        &2521  \\
  (4640/4650~\AA)        &&        &        &       &&       &       &       &&         &        &        &&       &        &       &&        &         &          &&         &        &        &&       &         &      \\
HeII (4686~\AA)	         && 27     & 13     &1154   &&27     &4      &807    && 9       & 7      &2286    &&15     &18      &960    &&65      &73       &1728      && 38      & 32     &1408    &&16     &11       &2091  \\
H$\beta$ (4861~\AA)      && 476    & 220    &1125   &&1270   &188    &1111   && 39      & 37     &1851    &&35     &43      &1358   &&95      &84       &2036      && 59      & 52     &1542    &&57     &47       &1557  \\
HeI (4922 ~\AA)          && 28     & 14     &1011   &&55     &8      &916    &&\nodata  &\nodata &\nodata &&2      &3       &727    &&10      &9        &1286      &&\nodata  &\nodata &\nodata &&5      &4        &1168  \\
HeI (5016~\AA)           && 23     & 13     &970    &&59     &11     &978    &&\nodata  &\nodata &\nodata &&\nodata&\nodata &\nodata&&\nodata &\nodata  &\nodata   &&\nodata  &\nodata &\nodata &&\nodata&\nodata  &\nodata\\
HeI (5875~\AA)	         && 131    & 89     &1014   &&3530   &70     &966    && 13      & 19     &1736    &&9      &16      &1481   &&28      &32       &1736      && 19      & 23     &1021    &&21     &25       &1652  \\
H$\alpha$ (6563~\AA)     && 510    & 341    &1024   &&1280   &287    &1000   && 46      & 94     &1582    &&40     &87      &1371   &&78      &110      &1599      && 62      & 81     &1188    &&63     &78       &1397  \\
HeI (6678~\AA)	         && 58     & 54     &1000   &&136    &32     &905    && 9       & 21     &2205    &&3      &8       &750    &&6       &37       &1438      && 6       & 9      &898     &&7      &8        &1123  \\
HeI (7065~\AA)           && 34     & 40     &1077   &&81     &32     &980    && \nodata &\nodata &\nodata &&\nodata&\nodata &\nodata&&\nodata &\nodata  &\nodata   &&\nodata  &\nodata &\nodata &&\nodata&\nodata  &\nodata\\
\hline
\end{tabular}                 	       
\tablecomments           	       
{Flux, EW, and FWHM are in the unit of 10$^{-16}$ erg cm$^{-2}$ s$^{-1}$, \AA, and km s$^{-1}$, respectively.}
\end{table*}
%-------------------------------------------------------------------

\subsubsection{Orbital-Phased Light Curve Variations and System's Geometry}
\label{sec:flc_RBS490}
We explored the phased-light curve variation of RBS 0490 during its binary motion. Light curves obtained from the TESS and R-band observations are folded using the time of first observation JD=2457305.572836 as the reference epoch. However, the sparse light curve spanning over several years in the ZTF-r band is not useful to produce meaningful phase-dependent light curve variations. Phase-folded light curves with phase bins of 0.0222 and 0.05 for TESS and R-band observations, respectively are shown in Figure \ref{fig:tess_opt_flc_RBS490}. Though, the R-band light curve is less extensive yet both TESS and R-band light curves appear to show almost consistent light curve variation. However, a careful inspection of the phased light curve from the TESS observations shows double-humped-like features which are separated by an orbital phase of 0.5, whereas the phased R-band light curve shows a plateau-like structure followed by a broad hump.

Using the mean empirical mass-period relation given by \citet{Smith98}, we have derived the mass and radius of the secondary as 0.10($\pm$0.02) $M_{\odot}$ and 0.16($\pm$0.01 $R_{\odot}$), respectively. Using an orbital period of 1.689 hr, we have estimated the mean density of secondary, $\overline{\rho}$=36.8 g cm$^{-3}$ \citep[see][for formulae]{Warner95}, which corresponds to a lower main-sequence star of spectral class M5.6 and an effective temperature of 2975 K \citep[see][]{Beuermann98, Knigge06}. Using the mass range of WD of 0.2$-$1.5 $M_{\odot}$ in CVs \citep{Zorotovic11}, the binary separation (a) is estimated to be in the range of (3.3$-$5.8) $\times$$10^{10}$ cm.

\subsubsection{Optical Spectroscopy}
\label{sec:spec_RBS490}
Optical spectra with a wealth of spectral features was observed for RBS 0490 during two different epochs of observations. Figure \ref{fig:optspec_RBS490} shows the optical spectra of RBS 0490 for epochs 22 December 2017 and 30 November 2018. Identification, flux, EW, and FWHM of the principal emission lines of RBS 0490 obtained through a single-Gaussian fitting are given in Table \ref{tab:opt_spec_RBS490_J0759}. Optical spectra of RBS 0490 exhibit a flat continuum and strong Balmer emission lines (from H$\alpha$-$H\zeta$), He I, and He II emission lines. The HeII $\lambda$4686 line is only modestly strong. The observed spectra resemble the spectra of \citet{Schwope02} and \citet{Thorstensen06} observed during epochs 1997, 1998, and 2002. Similar to \citet{Thorstensen06}, we observed all the spectral lines as single-peaked, whereas \citet{Schwope02} describes them as being double-peaked. The values of flux and EWs values are found to be variable during the present and previous epochs of observations \citep[see][and present work]{Schwope02, Thorstensen06}.

%%%%%%%%%%%%%%%%%%%%%%%%%%%%%%%%%%%%%%%%%%%%%%%%%%%%%%%%%%%%%%%%% SDSS J075939.79+191417.3%%%%%%%%%%%%%%%%%%%%%%%%%%%%%%%%%%%%%%%%%%%%%%%%%%%%%%%%%%%%%%%%%%%%%%%%%%

%==========================
\begin{figure*}[ht]
\centering
\includegraphics[width=165mm, height=50mm]{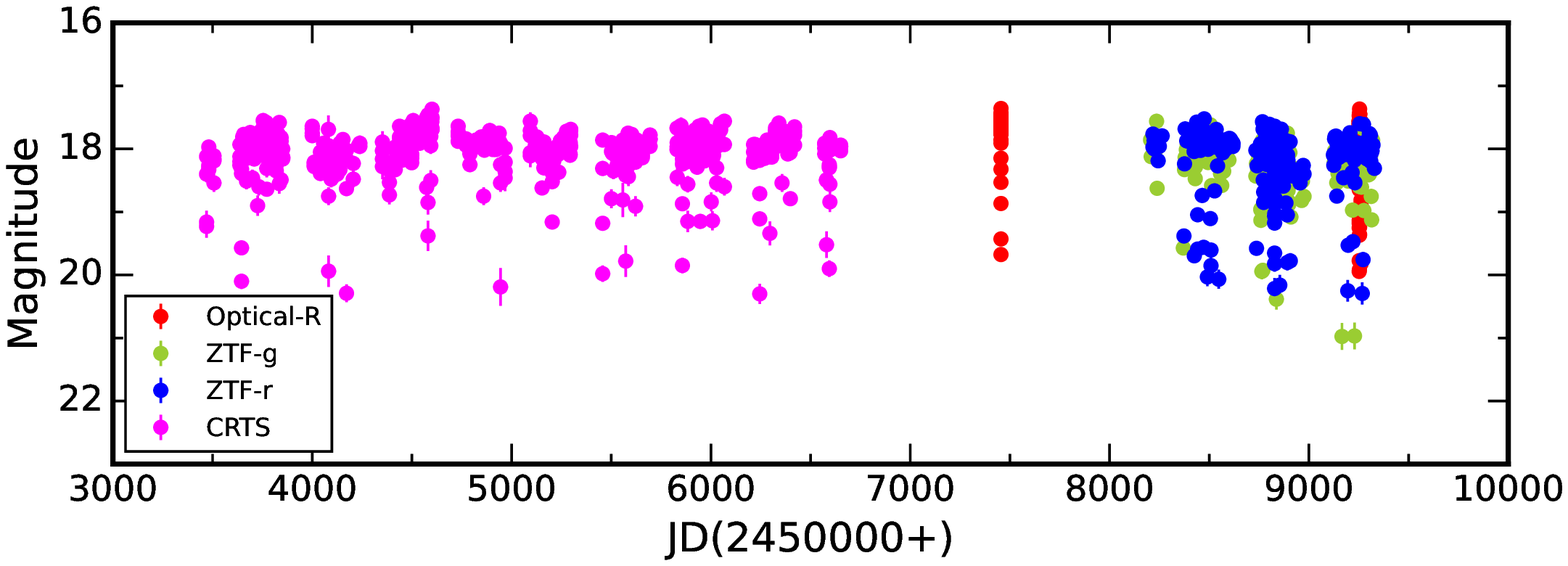}
\caption{Combined R-band, ZTF-r/g, and CRTS light curve variations of J0759. } 
\label{fig:longterm_lc_J0759}
\end{figure*}
%==========================
%==========================
\begin{figure*}[]ht
\centering
\includegraphics[width=160mm, height=90mm]{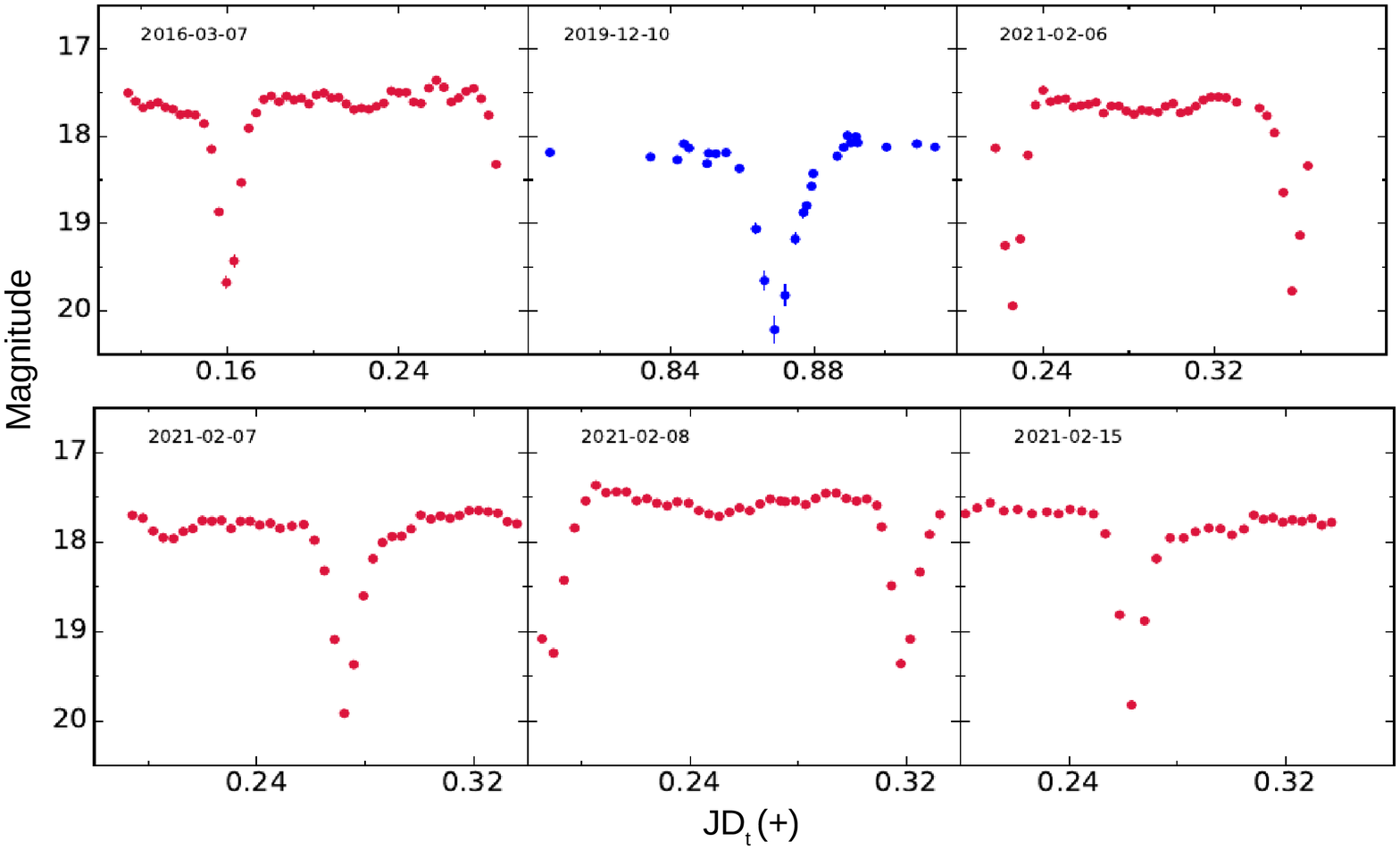}
\caption{R-band (red) and ZTF-r (blue) light curves of J0759, where the observation day (JD$_{t}$) is 2458827.00 for ZTF-r band light curve and for photometric observations it is given in Table \ref{tab:obslog}. The epoch of observations are mentioned at the top of each panel.} 
\label{fig:lc_J0759}
\end{figure*}
%==========================

\subsection{SDSS J075939.79+191417.3}\label{sec:ana_res_J0759}
\subsubsection{Light Curve Morphology and Power Spectra}\label{sec:lc_ps_J0759}
Figure \ref{fig:longterm_lc_J0759} shows complete CRTS, R-band, and ZTF-r/g light curves of J0759 where the variable nature of the source is clearly evident. R-band data of the epoch 2021 are found to be observed simultaneously with a few epochs of ZTF in the r and g bands. To probe the system deeply, we have further inspected the temporal properties of J0759 from the CRTS, R-band, and ZTF-r/g band observations separately. We have obtained a total of 17.76 hrs of photometric data from our observations carried over five nights. R-band light curves for five epochs of observations are shown in Figure \ref{fig:lc_J0759}. The light curves of J0759 exhibit a clear eclipse profile with an average dip of $\sim$ 2 mag and eclipse duration of $\sim$ 40 min. Successive eclipses spaced by $\sim$ 3.1 hr were observed on two of the five nights. Similar to the R-band light curve, the ZTF observations of J0759 also reveal the eclipse profiles in their light curves where the brightness drops by $\sim$ 2.0 mag as seen in the R-band light curve. A complete eclipse profile is seen in the ZTF-r band light curve for the epoch 10 December 2019 which is shown in the second panel of Figure \ref{fig:lc_J0759}.

\begin{figure}
\centering
\includegraphics[width=92mm, height=105mm]{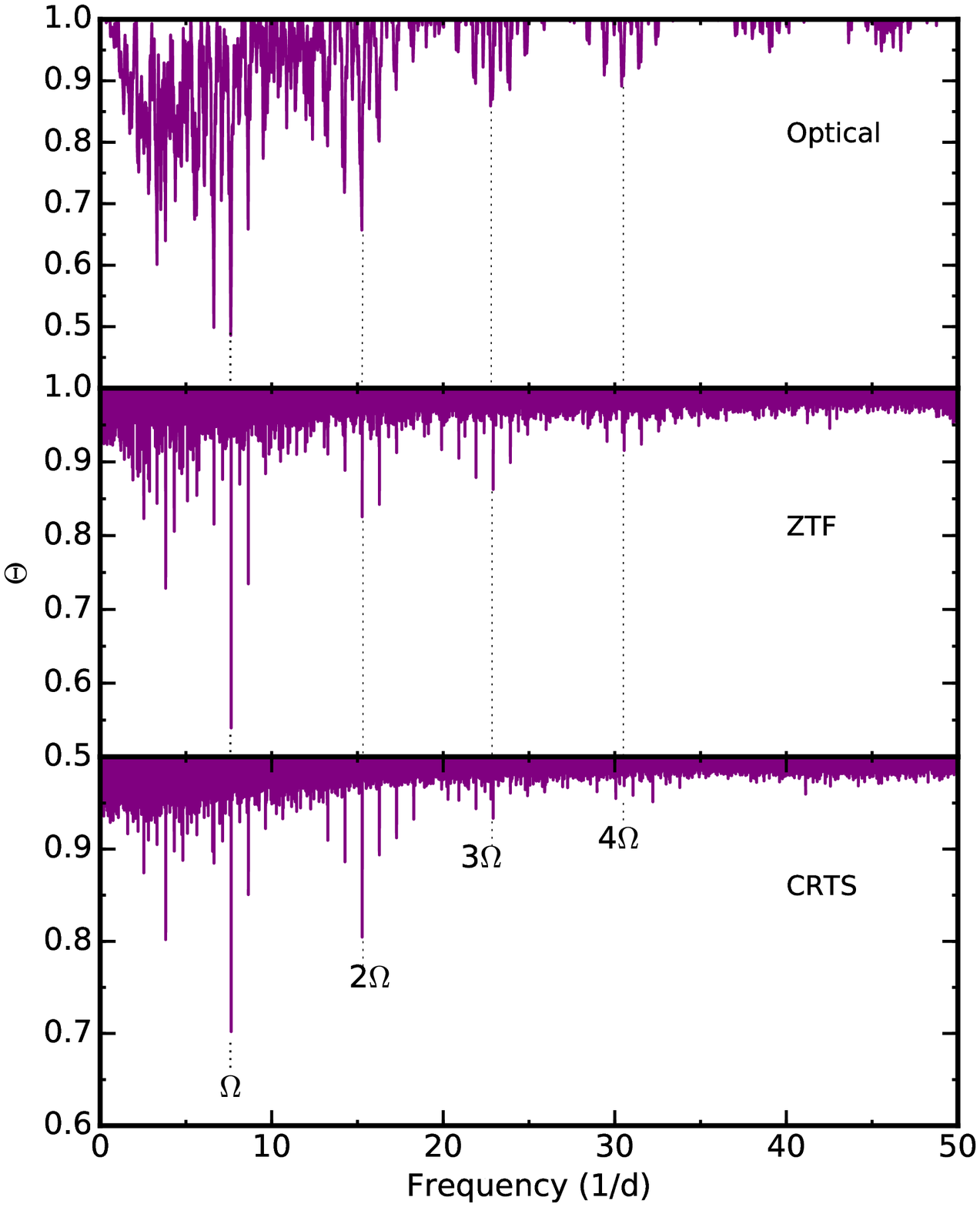}
\caption{Top to bottom panels show the PDM periodogram of J0759 as obtained from combined R-band photometric, ZTF, and CRTS observations, respectively. }
\label{fig:opt_ztf_crts_ps_J0759} 
\end{figure}

In order to find the periodicity from the eclipsed light curves of J0759, we have performed phase dispersion minimization \citep[PDM;][]{Stellingwerf78} algorithm to the combined photometric, ZTF, and CRTS observations. Because of the large data gap between epochs 2016 and 2021, we have used combined photometric data observed during the epoch 2021 for periodogram analysis. The PDM power spectra of R-band, ZTF, and CRTS data are shown in Figure \ref{fig:opt_ztf_crts_ps_J0759}. Various prominent peaks are found in the PDM power spectra. The first prominent peak in the PDM periodogram of photometric, ZTF, and CRTS data corresponds to the orbital period of $\sim$ 3.1 hr and perfectly matches with the time-span where primary eclipse completes its one cycle (see Figure \ref{fig:lc_J0759}). Other peaks are also observed in the PDM power spectra which correspond to the harmonics of the orbital frequencies.

%==========================
%==========================

%-------------------------------------------------------------------
\begin{table}[ht]
\normalsize
\setlength{\tabcolsep}{0.05in}
\centering 
\caption{Times of minima for J0759 from R-band and ZTF observations.\label{tab:eclipsetime_0759}}
\begin{tabular}{ccccccccc}
 \hline 
{\textbf {Date of Obs.}} & {\textbf {Eclipse Minima (JD)}}      &  {\textbf {Cycle}}           \\
\hline
2016 Mar 07   & 2457455.16058506$\pm$0.00014480&\colhead{0}         \\  
2019 Dec 10   & 2458827.86967765$\pm$0.00020655&\colhead{10484.00} \\
2021 Feb 06   & 2459252.22568635$\pm$0.00015772&\colhead{13725.00}  \\ 
              & 2459252.35599537$\pm$0.00014890&\colhead{13725.99}  \\
2021 Feb 07   & 2459253.27295816$\pm$0.00021381&\colhead{13732.99}  \\
2021 Feb 08   & 2459254.31926978$\pm$0.00025254&\colhead{13740.99} \\ 
2021 Feb 15   & 2459261.26253060$\pm$0.00025446&\colhead{13794.01}  \\[1.0ex]
\hline\\
\end{tabular}                 	       
\end{table}
%-------------------------------------------------------------------

\subsubsection{Orbital-Phased Light Curve Variations and System's Geometry}
\label{sec:flc_J0759}
\begin{figure*}[ht]
\centering
\includegraphics[width=160mm, height=80mm]{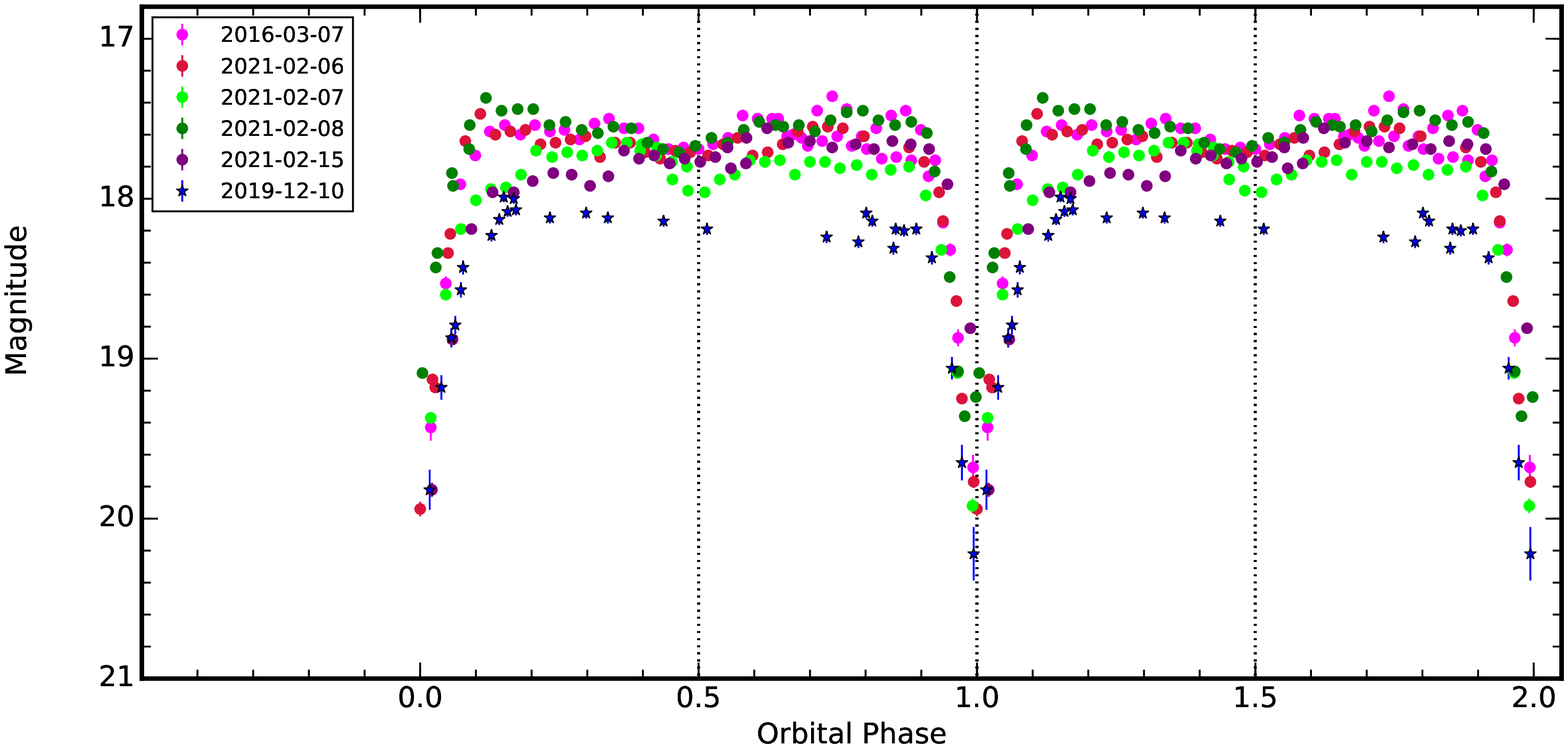}
  \caption{Orbital-phase-folded light curves of J0759 in R-and ZTF-r bands.}
  \label{fig:flc_J0759}
\end{figure*}

We have determined a total of 7 mid-eclipse timings (see Table \ref{tab:eclipsetime_0759}) by fitting a Gaussian function to the bottom part of eclipses. Only those epochs of observations were used for which complete eclipse profiles were observed. A linear least-squares fit to these eclipse timings yields 

\begin{equation}                         
T_0 = JD~2457455.16054(45) \pm 0.13093372(4) E,
\label{eq:eph_J0759}
\end{equation}                         

\noindent
corresponding to $P_\Omega$ of 3.14240928$\pm$0.00000096 hr. We have also folded the ZTF-r band light curve of the epoch 10 December 2019 and R-band light curves using ephemeris as derived in equation \ref{eq:eph_J0759}. Figure \ref{fig:flc_J0759} shows the orbital-phase-folded light curves of J0759. Each light curve of J0759 reveals a deep eclipse profile with variable brightness. We have also found evidence of a secondary eclipse near phase 0.5 during all epochs of observations with approximate depths in the range of 0.14-0.3 mag. We could not see any eclipse around phase 0.5 in the ZTF-r band light curve due to its sparse phase coverage.

Using the eclipse width at half depth ($\Delta \phi$), we derive the orbital inclination of the binary system by using the equation given by \citep{Eggleton83},
\begin{equation}
\left( \frac{R_2}{a} \right)^2  = \sin^2(\pi \Delta \phi) + \cos^2(\pi \Delta \phi) \cos^2i,
\label{eq:R2/a}
\end{equation}
where $R_2/a$ depends only on the mass ratio as 
\begin{equation}
\frac{R_2}{a} =  \frac{Cq^{2/3}}{Dq^{2/3}+\ln(1+q^{1/3})}.
\label{eq:R2/a_1}
\end{equation}  
\noindent
The coefficients $C$ and $D$ are given by \citet{Eggleton83} as 0.49 and 0.6, respectively for a spherical shape of Roche lobe. However, in the case of CVs, the Roche lobe of secondary is not spherical, having the largest size as ``seen'' from the white dwarf and the smallest in the polar direction. Therefore, we have used the coefficients $C=0.4990$ and $D=0.5053$ \citep{Andronov14}. Further, using the same approach as described in section \ref{sec:flc_RBS490}, the mass and radius of the secondary are estimated as 0.28($\pm$0.01) $M_{\odot}$ and 0.33($\pm$0.01 $R_{\odot}$), respectively. Assuming the minimum and maximum values of WD in CVs of 0.2 $M_\odot$ and 1.5 $M_\odot$, the limiting values of $q$ are estimated as 1.4 and 0.19, respectively. For each consecutive epochs of R-band observations (see Table \ref{tab:obslog}) and ZTF-r band observations of the epoch 10 December 2019, the eclipse width at half depth ($\Delta \phi$) was estimated as 0.081$\pm$0.003, 0.077$\pm$0.004, 0.078$\pm$0.005, 0.082$\pm$0.004, 0.068$\pm$0.005, and 0.087$\pm$0.004, respectively. Adopting the average values of $\Delta \phi$ and  $q$ of 0.079$\pm$0.002 and 0.35 (assuming the average value of the mass of WD in CVs as 0.8 $M_\odot$; \cite{Zorotovic11}), respectively, the orbital inclination ($i$) is estimated to be $\sim$ 78\hbox {$^\circ$}. This estimated value of $i$ is in good agreement with the values derived in terms of the graphical form of the relationship between $\Delta \phi$, $i$, and q for Roche geometry in \cite{Horne85}. Considering the limiting values of WD masses in CVs, the binary separation is estimated to be in the range of (5.9-9.2)$\times$10$^{10}$ cm. The mean density of secondary $\overline{\rho}$ is estimated to be 10.8 g cm$^{-3}$, which infers the spectral type and effective temperature of secondary as M4.2 and 3292 K, respectively. 

The total duration of an eclipse is approximately 40 min. This measured value of $\sim$ 40 min for the entire eclipse duration is too long to represent an eclipse of just the primary star and can also be used to determine the size of the eclipsed light-producing source. Thus, we have determined the radius of the eclipsed region ($R$) by using the following equation of \cite{Bailey90}
\begin{equation}
 R =  \pi ~a ~\sqrt{(1-\alpha^2)} ~\Delta \phi_{ie} ,
\label{eq:eq5}
\end{equation}
\noindent 
where $\Delta \phi_{ie}$ is ingress/egress duration,  $\alpha$=$cos~i$/$cos~i_{lim}$, $i$ is angle of inclination, and $i_{lim}$ is a limiting angle of inclination for which eclipse half-width at half depth reaches zero. The average value of ingress/egress duration ($\Delta \phi_{ie}$) is derived to be $0.14\pm0.06$, where the error on $\Delta \phi_{ie}$ is the standard deviation of different measurements. Assuming the angle of inclination, $i \sim 78^\circ$, average value of $q$=0.35, and using $i_{lim}$ from the Table 1 of \cite{Bailey90}, the value of $R$ is estimated to be $\sim$ 33 $R_{WD}$, where $R_{WD}$ is calculated using the mass-radius relation given by \citet{Nauenberg72}.

%==========================
%==========================

\subsubsection{Optical Spectroscopy}
\label{sec:spec_J0759}
\begin{figure*}
\centering
\includegraphics[width=90mm,height=110mm]{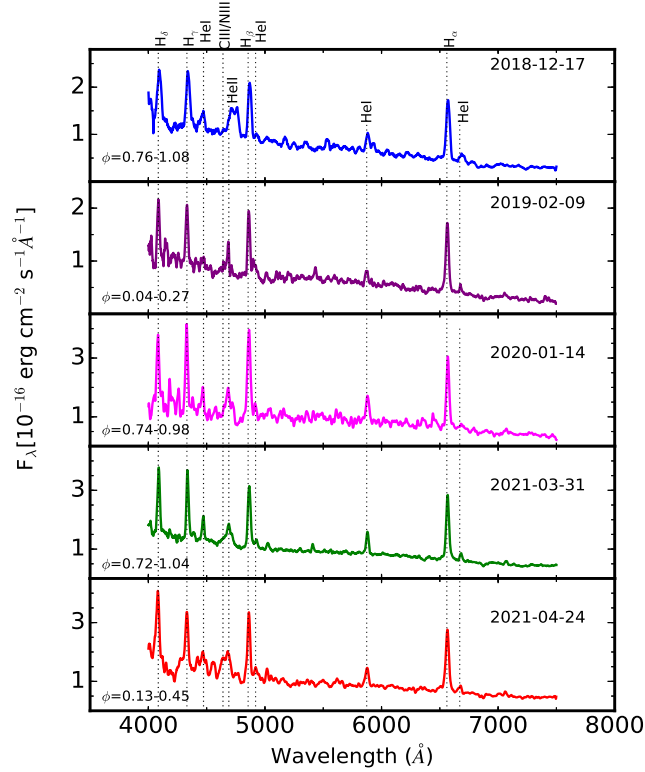}
\caption{Optical spectra of J0759 for five epochs of observations. The date of observation and corresponding orbital phase are mentioned in each panel of the spectrum.}
\label{fig:optspec_J07}
\end{figure*}

Optical spectra of J0759 obtained for five epochs of observations are shown in Figure \ref{fig:optspec_J07}. The orbital phase for every epoch of observations was derived using mid-time of exposure and the equation \ref{eq:eph_J0759}. In each spectrum, we have observed strong Balmer emission lines from H$\alpha$ to H$\delta$, along with the He I and weak He II ($\lambda$4686 \AA) lines. We have also found the weak Bowen fluorescence CIII/NIII emission lines in the spectra of J0759. These emission lines are found to be variable from one epoch of observation to another. The current optical spectra of J0759 resemble the earlier observed optical spectrum from the SDSS as obtained by \cite{Szkody06} during the epoch 06 November 2004. The SDSS flux of the emission lines H$\alpha$, H$\beta$, and He I ($\lambda$ 4471 \AA) seems to be consistent with the observed flux for the epoch 09 February 2019, however, the strength of emission line He II ($\lambda$4686 \AA) was relatively weak in the SDSS spectrum \citep{Szkody06}.

%==========================
%==========================

%%%%%%%%%%%%%%%%%%%%%%%%%%%%%%%%%%%%%%%%%%%%%%%%%%%%%%%%%%%%%%%%%%%%% Discussion %%%%%%%%%%%%%%%%%%%%%%%%%%%%%%%%%%%%%%%%%%%%%%%%%%%%%%%%%%%%%%%%%%%%%%%%%%%%%%%%%%%

\section{Discussion}
\label{sec:diss}
We have carried out detailed analyses of two CVs $-$ RBS 0490 and J0759, using optical photometry and spectroscopy. We discover a photometric period of 1.689$\pm$0.001 hr for RBS 0490, which is probably its orbital period. For J0759, our eclipse timings confirm and refine the previously determined period of 3.14-hr. The orbital period of RBS 0490 and J0759 place them below and above the period gap of orbital period distribution of CVs, respectively.

The absolute value of G-band magnitude is approximately +9.9 mag for RBS 0490 which appears to be similar to that for the magnetic systems of AM Her types. Further, the photometric light curve variations in RBS 0490 also look similar to a magnetic system, possibly polars, and these variations are unusual for an ordinary nova-like or a dwarf nova. These light curve variations reveal interesting one broad hump and then a second hump or a plateau-like structure. The main broad hump can be associated with the main accretion region. However, the observed second hump or plateau-like structure indicates that the minimum might be filled up by the emission of an independent second accretion region or may indicate accretion at a second fainter pole. The accretion onto a second pole has been seen in a few polars, e.g. in DP Leo \citep{Cropper90}, VV Pup \citep{Wickramasinghe89}, UZ For \citep{Schwope90}, and QS Tel \citep{Schwope95}, etc. In the majority of the cases, the activity from the second pole is at least one order of magnitude lower as compared to the primary one. The polarimetric observations of RBS 0490 can help to confirm the two-pole accretion in the future.

R-band light curves of J0759 exhibit deep eclipses with an average magnitude of the depth of $\sim$ 2.0 mag. The observed eclipse resembles the disc rather than the highly structured eclipses often seen in AM Her stars. The absolute G-band magnitude implied by the Gaia parallax is near +6.5, significantly brighter than most magnetic systems. Further, the large value of eclipse radius indicates that the eclipse observed in the light curves is not solely due to the occultation of WD by secondary. In general, the dimension of the optical emission sites is thought to be $<$ 0.1 $R_{WD}$ \citep{Bailey91}. Therefore, the derived large value of $R$ indicates that the eclipse feature might be observed due to the occultation of the accretion disc and bright spot. 

The optical spectra of J0759 show strong Balmer emission lines, He I, and He II along with conspicuous features of weak He II ($\lambda$4686 \AA) and Bowen fluorescence CIII/NIII emission lines. The weakness of HeII $\lambda$4686 and the Bowen lines suggest that J0759 is not a magnetic CV \citep{Warner95}. However, this is not a sufficient criterion for a CV classification. More specifically, the ratio of EW[He II ($\lambda$4686)/$H\beta$] and EW of $H\beta$ emission line provides a criterion to separate magnetic CVs from non-magnetic CVs \citep{Silber92}. According to Silber's criteria $-$ magnetic CVs are characterized by the ratio EW[He II ($\lambda$4686)/$H\beta$] $\geq$ 0.4 and $H\beta$ $\geq$ 20 \AA. For J0759, we have detected the $H\beta$ $\geq$ 20 \AA~ for all observations, however, the EW ratio of He II ($\lambda$4686) to $H\beta$ is not detected more than 0.4 for all epochs of observations. For the epochs 17 December 2018 and 24 April 2021, we have observed this ratio is close to 0.2, which is not quite up to the values for magnetic CVs. The Balmer decrement can also be used in order to distinguish the magnetic and non-magnetic nature of the system. The inverted Balmer decrement is possibly present in the low-state spectra of the AM Her type of magnetic CVs. However, a flat Balmer decrement, implying that the Balmer emission lines are optically thick, has been observed in most of the non-magnetic CVs and also in intermediate polars \citep{Warner95}. For J0759, a flat Balmer decrement is observed in the most epoch of observations. The cyclotron hump features are also not detected in the optical spectra of J0759. Thus, it appears that the system J0759 follows the majority of the criteria of being non-magnetic but further observations are needed for definitive classification.

In the case of RBS 0490, optical spectra also consist of strong emission lines of Balmer series and He I together with the weak emission line of He II ($\lambda$4686). The presence of single-peaked strong Balmer emission lines and the large equivalent width of the H$\beta$ emission line may imply a magnetic CV classification for this system. However, other possible factors like weak He II ($\lambda$4686 \AA), broad Balmer emission lines, the observed flat Balmer decrement, the ratio of EW[He II ($\lambda$4686)/$H\beta$] $\sim$ 0.1 and 0.02 (for epochs 22 December 2017 and 30 November 2018), and the non-detection of cyclotron hump in the optical spectrum either weakens its chance to be magnetic or may imply the low magnetic strength of the WD. A low magnetic ﬁeld strength of the WD was also proposed by \cite{Harrison15} from the combined WISE and JHK observations of RBS 0490.

%%%%%%%%%%%%%%%%%%%%%%%%%%%%%%%%%%%%%%%%%%%%%%%%%%%%%%%%%%%%%%%%%%%%% Conclusions%%%%%%%%%%%%%%%%%%%%%%%%%%%%%%%%%%%%%%%%%%%%%%%%%%%%%%%%%%%%%%%%%%%%%%%%%%%%%%%%%%%

\section{Conclusions}
\label{sec:conc}
We have presented the optical photometric and spectroscopic observations of RBS 0490 and J0759 which leads us to the following conclusions:
\begin{itemize}
\item RBS 0490 has been found to vary with a period of 1.689$\pm$0.001 hr, which was not evident in earlier studies and can be interpreted as its orbital period. The probable orbital period places it below the period gap of the orbital period distribution of CVs. The variability observed in RBS 0490 seems to favour the magnetic systems, possibly polars, and provide evidence of the emission from an independent second accretion region or a second fainter pole. However, the characteristic features seen in the optical spectra of RBS 0490 either weaken its chance to be magnetic or may imply the low magnetic strength of the WD.
\item Our eclipse photometry confirms and refines the $\sim$ 3.14 hr period of J0759, placing it longward of the period gap. The eclipses are consistent with a disc and an orbital inclination of $\sim$ 78\hbox {$^\circ$}. The detection of strong Balmer emission lines along with the weak high ionization emission lines hint towards the non-magnetic nature of J0759.  
\end{itemize}

%%%%%%%%%%%%%%%%%%%%%%%%%%%%%%%%%%%%%%%%%%%%%%%%%%%%%%%%%%%%%%%%%%%%%% Acknowledgements%%%%%%%%%%%%%%%%%%%%%%%%%%%%%%%%%%%%%%%%%%%%%%%%%%%%%

\section*{Acknowledgements}

We acknowledge the referee for useful comments and suggestions that improved the manuscript considerably. This work is supported by the National Key Research and Development Program of China (Grants No. 2021YFA0718500, 2021YFA0718503), the NSFC (12133007, U1838103, 11622326), and the Fundamental Research Funds for the Central Universities (No. 2042021kf0224). This research includes data collected with the TESS mission, obtained from the MAST data archive at the Space Telescope Science Institute (STScI). Funding for the TESS mission is provided by the NASA Explorer Program. STScI is operated by the Association of Universities for Research in Astronomy, Inc., under NASA contract NAS 5–26555. Based on observations obtained with the Samuel Oschin 48-inch and the 60-inch Telescope at the Palomar Observatory as part of the Zwicky Transient Facility project. ZTF is supported by the National Science Foundation under Grant No. AST-1440341 and AST-2034437 and a collaboration including Caltech, IPAC, the Weizmann Institute for Science, the Oskar Klein Center at Stockholm University, the University of Maryland, the University of Washington, Deutsches Elektronen-Synchrotron and Humboldt University, Los Alamos National Laboratories, the TANGO Consortium of Taiwan, the University of Wisconsin at Milwaukee, Trinity College Dublin, Lawrence Livermore National Laboratories, Lawrence Berkeley National Laboratories, and IN2P3, France. Operations are conducted by COO, IPAC, and UW. One of the authors Ashish Raj acknowledges the Research Associate Fellowship with order no. 03(1428)/18/EMR-II under Council of Scientific and Industrial Research (CSIR). The observing staff and observing assistants of 1-m and 2-m class telescopes are deeply acknowledged for their support during optical observations.

%%%%%%%%%%%%%%%%%%%% %%%%%%%%%%%%%%%%%%%%%%%%%%%%%%%%%%%%%%%%%%%%%% REFERENCES %%%%%%%%%%%%%%%%%%%%%%%%%%%%%%%%%%%%%%%%%%%%%%%%%%%%%%%%%%%%%

\bibliographystyle{aasjournal}
\bibliography{ref}

%%%%%%%%%%%%%%%%%%%%%%%%%%%%%%%%%%%%%%%%%%%%%%%%%%%%%%%%%%%%%%%% End %%%%%%%%%%%%%%%%%%%%%%%%%%%%%%%%%%%%%%%%%%%%%%%%%%%%%%%%%%%%%%%%%%%%%%%%%%%%%%%%%%%%%%%%%%%%%%%

\end{document}